\documentclass[aip,jap,reprint]{revtex4-1}

\usepackage{graphicx}
\usepackage{hyperref}
\usepackage{hypcap}

\hypersetup{ 
colorlinks=true,
linkcolor=blue,
citecolor=blue
} 

\newcommand{\mytilde}{\raise.17ex\hbox{$\scriptstyle\mathtt{\sim}$}}

\begin{document}

\title{Structural defects induced by Fe-ion implantation in TiO$_2$}

\author{B. Leedahl}
\email[]{bdl816@mail.usask.ca}
\affiliation{Department of Physics and Engineering Physics, University of Saskatchewan, 116 Science Place, Saskatoon, Saskatchewan, Canada, S7N 5E2}

\author{D. A. Zatsepin}
\affiliation{Ural Federal University, 19 Mira Str., 620002 Yekaterinburg, Russia}

\author{D. W. Boukhvalov}
\affiliation{School of Computational Sciences, Korea Institute for Advanced Study (KIAS) Hoegiro 87, Dongdaemun-Gu, Seoul, 130-722, Korean Republic}

\author{R. J. Green}
\affiliation{Department of Physics and Engineering Physics, University of Saskatchewan, 116 Science Place, Saskatoon, Saskatchewan, Canada, S7N 5E2}

\author{J. A. McLeod}
\affiliation{Department of Physics and Engineering Physics, University of Saskatchewan, 116 Science Place, Saskatoon, Saskatchewan, Canada, S7N 5E2}
\altaffiliation{Current Address: College of Nano Science and Technology, Soochow University, 199 Ren-Ai Rd., Suzhou Industrial Park, Suzhou, Jiangsu, 215123, China}

\author{S. S. Kim}
\affiliation{School of Materials Science and Engineering, Inha University,Incheon 402-751, 
Republic of Korea}

\author{E. Z. Kurmaev}
\affiliation{Institute of Metal Physics, Russian Academy of Sciences-Ural Division, 620990 Yekaterinburg, Russia}

\author{I. S. Zhidkov}
\affiliation{Ural Federal University, 19 Mira Str., 620002 Yekaterinburg, Russia}

\author{N. V. Gavrilov}
\affiliation{Institute of Electrophysics, Russian Academy of Sciences-Ural Division, 620016 Yekaterinburg, Russia}

\author{S. O. Cholakh}
\affiliation{Ural Federal University, 19 Mira Str., 620002 Yekaterinburg, Russia}

\author{A. Moewes}
\affiliation{Department of Physics and Engineering Physics, University of Saskatchewan, 116 Science Place, Saskatoon, Saskatchewan, Canada, S7N 5E2}

\date{\today}

\begin{abstract}
X-ray photoelectron spectroscopy (XPS) and resonant x-ray emission spectroscopy (RXES) measurements of pellet and thin film forms of TiO$_2$ with implanted Fe ions are presented and discussed. The findings indicate that Fe-implantation in a TiO$_2$ pellet sample induces heterovalent cation substitution (Fe$^{2+}\rightarrow$ Ti$^{4+}$) beneath the surface region. But in thin film samples,  the clustering of Fe atoms is primarily detected. In addition to this, significant amounts of secondary phases of Fe$^{3+}$ are detected on the surface of all doped samples due to oxygen exposure. These experimental findings are compared with density functional theory (DFT) calculations of formation energies for different configurations of structural defects in the implanted TiO$_2$:Fe system. According to our calculations, the clustering of Fe-atoms in TiO$_2$:Fe thin films can be attributed to the formation of combined substitutional and interstitial defects. Further, the differences due to Fe doping in pellet and thin film samples can ultimately be attributed to different surface to volume ratios.
\end{abstract}


\maketitle

\section{Introduction}
Enormous efforts have been put forth in the last several years to study TiO$_2$-based materials; these materials have been shown to display many promising technological applications in a wide variety of fields. This includes areas such as photovoltaics,\cite{gratzel2001} photocatalysis,\cite{hagfeldt1995,linsebigler1995,millis1997} photo/electrochromics,\cite{russo2010} and spintronics.\cite{macdonald2005, Chang2006} The electronic properties of the TiO$_2$ host material, and therefore the specific technological application as well, depend on the modifications to the sample (for example, by ordinary chemical doping, precise cation-anion site substitutions, etc.). In addition to this, the interactions between the TiO$_2$-based materials and their surrounding environment play an active role in the properties of such materials, and should therefore also be considered when studying these materials.\cite{chen2007}

Among the aforementioned modifications to the TiO$_2$ matrix, cation doping with 3$d$-transition metals is of particular interest. This is because 3$d$ transition metal doping is often linked with the appearance of ferromagnetism in dilute magnetic semiconductors, since ferromagnetism can be induced by the exchange interaction of magnetic $3d$-ions mediated by carriers.\cite{ohtsuki2011, Chang2009} A second point of interest is that by filling the mid-gap states in TiO$_2$ with $d^n$-states, the band gap ($\sim$3.03 eV for rutile TiO$_2$),\cite{Pascual1978} which is too large for absorption in the visible part of solar spectrum, will be reduced. The system would then be more viable for established photocatalysis use in the visible region of the electromagnetic spectrum.\cite{hashimoto2005} 

Doping of TiO$_2$ with Fe, Co and Ni occurs only by heterovalent substitution, this is because these 3$d$-elements do not easily maintain a 4+ oxidation state. This induces the formation of different structural defects (vacancies, interstitials, precipitates), which can affect the electronic structure, and hence, the magnetic and electrical properties of doped materials. 

In the present paper we have studied the local structure of Fe impurity atoms in TiO$_2$ pellet and thin films using x-ray photoelectron spectroscopy (XPS) and resonant x-ray emission spectroscopy (RXES). Based on these measurements, the structural models of TiO$_2$:Fe are discussed and compared with our DFT-calculations. 

\section{Experimental and Calculation Details}
\subsection{Sample Preparation}
TiO$_2$ coating sols were prepared by a sol-gel chemical process, where titanium-isopropoxide, nitric acid, and anhydrous ethanol were used as the precursor, catalyst, and solvent, respectively. Deionized water was also supplied for the hydrolysis of TIPP and all of the chemicals were used as received without any further purification. Using the prepared coating sols, TiO$_2$ films were deposited on Si wafers (100) by a dip-coating process.\cite{katoch2012} The withdrawal rate of the substrate was 4 mm/s. The as-prepared films were dried at room temperature and then kept in an oven at 60$^\circ$C for 1 day to remove the remaining solvents completely, and finally they were annealed at 100$^\circ$C for 2 hours. The obtained films were $\sim$200 nm thick and characterized by field emission scanning electron microscopy and atomic force microscopy.

Samples of ceramic TiO$_2$ powder were obtained by electrical explosion of wires,\cite{kotov2003} they were made in molds of 15 mm diameter at $7 \times 10^4$ N force. They were then sintered for one hour at a temperature of 1040$^\circ$C. The final dimensions when compact were on average: \mbox{12.8 mm} in diameter, \mbox{1.8 mm} in height, and a density of \mbox{4.25 g/cm$^3$}. The phase composition was verified by an x-ray diffractometer (XRD), and the compact material was found to be nearly all single phase rutile (99.85\%). The parameters of its tetragonal lattice were: \mbox{a = b = 4.592 \AA, c = 2.960 \AA}, and the average crystallite size was determined to be \mbox{$>$ 200 nm}. This sample will be referred to herein as the ``pellet'' sample.

\subsection{Ion Implantation}
The implantation of Fe ions in pellet and thin film TiO$_2$ samples was carried out in vacuum, the chamber was evacuated to a residual pressure of \mbox{$3 \times 10^{-3}$ Pa}. An ion beam with an energy of 30 keV was then generated by the source based on a cathodic vacuum arc. The arc was initiated with an auxiliary discharge in an argon atmosphere, by doing this the gas pressure in the chamber increased to \mbox{$1.5 \times 10^{-2}$ Pa}. The processing was carried out in a pulsed mode with a repetition rate of 25 Hz and a pulse duration of 0.4 ms with a pulse current density of \mbox{0.7 mA/cm$^2$}. The duration of exposure for which the fluence reached \mbox{$1 \times 10^{17}$ cm$^{-2}$} was 38 minutes. The samples were mounted on a massive water-cooled collector in order to prevent overheating. The initial temperature of the samples prior to irradiation was 20$^\circ$C. After implantation, the samples were cooled under vacuum for 20 minutes. Stopping range of ions in matter (SRIM) simulations were performed to determine the approximate distribution and concentration of implanted ions.\cite{SRIM} The average concentration of Fe ions was found to be $\sim$23\% (by atomic \%) to a maximum depth of $\sim$45nm.

\subsection{XPS Measurements}
XPS core-level and valence-band measurements were made using a PHI XPS Versaprobe 5000 spectrometer (ULVAC-Physical Electronics, USA) based on the classic x-ray optic scheme with a hemispherical quartz monochromator and an energy analyzer working in the range of binding energies from 0 to 1500 eV. This system uses electrostatic focusing and magnetic screening to achieve an energy resolution of \mbox{$\Delta$E $\le$ 0.5 eV} for Al $K\alpha$ excitations \mbox{(1486.6 eV)}.  All samples under study were introduced to vacuum (10$^{-7}$ Pa) for 24 hours prior to measurement, and only samples whose surfaces were free from micro-impurities were measured and reported herein. The XPS spectra were recorded using Al $K\alpha$ x-ray emission; the spot size was 100 $\mu$m, and the x-ray power load on the sample was kept below 25 watts. Typical signal to noise ratios were above 10000:3. The spectra were processed using ULVAC-PHI MultiPak Software 9.3 and the residual background was removed using the Tougaard method.\cite{tougaard1987} XPS spectra were calibrated using a reference energy of 285.0 eV for the carbon 1$s$ level.\cite{moulder1992}

\subsection{XES and XAS Measurements}
The x-ray emission spectroscopy (XES) and x-ray absorption spectroscopy (XAS) measurements taken at the Fe $L$-edge were performed using Beamline 8.0.1 \cite{jia1995} at the Advanced Light Source (ALS) at the Lawrence Berkeley National Laboratory. The beamline uses a Rowland circle geometry grating spectrometer with spherical gratings. The photons emitted from the sample were detected at an angle of 90$^\circ$ with respect to the incident photons, and the incident photons were 30$^\circ$ to the sample surface normal with a linear polarization in the horizontal scattering plane. All of the experiments were performed in a vacuum chamber at \mbox{$\sim$$10^{-5}$ Pa.} The XAS resolving power (E$/\Delta$E) was $\sim$2000, while the XES resolving power was $\sim$1000.

The Fe $L$-edge XAS spectra were calibrated using a reference energy of 708.4 eV for the first peak in the $L_3$ absorption edge and a reference splitting of 13.5 eV between the $L_3$ and $L_2$ absorption lines of metallic iron; the XES spectra were then calibrated with respect to the elastic scattering peaks from incident x-rays with energies resonant with the $L_2$ and $L_3$ absorption lines. The O $K$-edge XES and XAS spectra were calibrated using a reference energy of 526.0 eV and 532.7 eV for the O $K$ emission line and absorption edge of Bi$_4$Ge$_3$O$_{12}$, respectively. Inverse partial fluorescence yield (IPFY)\cite{Regier2011} measurements were made using the SGM beamline at the Canadian Light Source.\cite{Regier2007} In IPFY, the edge of interest is resonantly excited over a a range of energies, but regular partial fluorescence yield (PFY) measurements are taken with the detector at some other lower energy edge of a \emph{different} element in the sample. As the resonantly excited (Fe $L_{2,3}$ is the edge of interest herein) edge begins absorbing photons, it reduces the amount absorbed at the lower energy edge (O $K$-edge herein), when inverted, this lower energy PFY spectra is proportional to the true x-ray absorption coefficient, but free of saturation and self-absorption effects.


\subsection{DFT Calculations}
The density-functional theory (DFT) calculations were performed using the SIESTA pseudopotential code, \cite{strange2008} as has previously been utilized with success for related studies of impurities in semiconductors.\cite{chang2007} All calculations were made using the Perdew-Burke-Ernzerhof variant of the generalized gradient approximation (GGA-PBE)\cite{perdew1996} for the exchange-correlation potential. A full optimization of the atomic positions was done, during which the electronic ground state was consistently found using norm-conserving pseudopotentials for the cores and a \mbox{double-$\xi$} plus polarization basis of localized orbitals for Fe, Ti, and O. Optimizations of the force and total energy were performed with an accuracy of \mbox{0.04 eV/\AA} and \mbox{1.0 meV}, respectively. For the atomic structure calculations, a Ti pseudopotential was employed with Ti $3d$ electrons treated as localized core states. The calculations of the formation energies ($E_{form}$) were performed using the standard method described in detail in Ref. \onlinecite{chang2007}. As a host for the studied defects, a TiO$_2$ supercell consisting of 96 atoms was used. Taking into account our previous modelling of transition metal impurities in semiconductors,\cite{chang2007} we have calculated various combinations of substitutional (S) and interstitial (I) Fe impurities.

GGA-PBE was chosen as opposed to more complex approaches such as using a hybrid functional or on-site Hubbard U because GGA-PBE is computationally much simpler than other approaches, and GGA-PBE adequately reproduces the electronic structure of TM-doped DMS systems.\cite{chang2007} Indeed, in past studies the more complex approaches provided virtually identical results in terms of defect structures and exchange interactions.\cite{Garcia2012, Sato2010}

\begin{figure}
\includegraphics[width=3.375in]{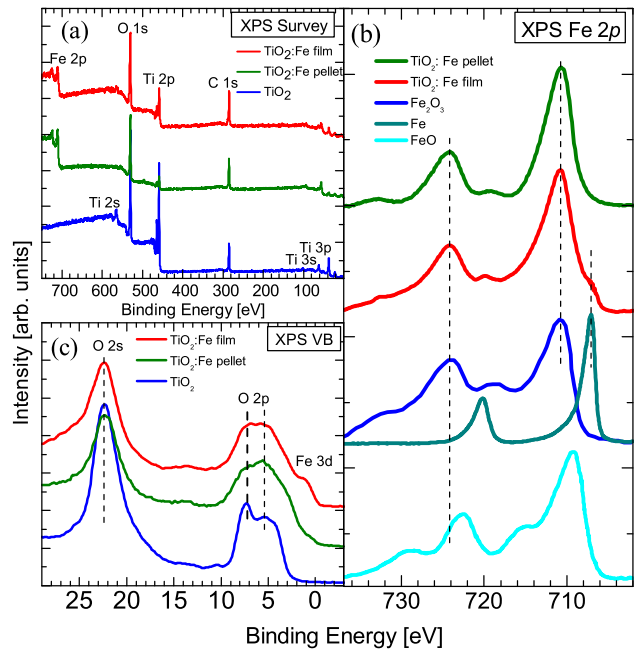}
\caption{(Color online) XPS data. (a) XPS Survery of pure and Fe-implanted TiO$_2$. The Fe $2p$ signal is strong in doped samples while the carbon content is relatively low, there is also no indication of impurites other than Fe. (b) Fe 2$p$ XPS spectra of doped and reference samples are shown. Both doped samples show a strong similarity to Fe$_2$O$_3$ in both pellet and thin film TiO$_2$ samples, indicating the presence of primarily Fe$^{3+}$ on the surface. In addition to this, a clear Fe metal signal can be seen in the thin film sample due to the clustering of Fe atoms. (c) Valence band spectra show that Fe doping introduces Fe $3d$ states near the Fermi level as compared to pure TiO2$_2$, indicated a reduction in the band gap upon doping with Fe.}
\label{Fig:XPS}
\end{figure}

\section{Results and Discussion}
By using core level XPS to probe the occupied density of states (DOS), the local environment of the absorbing atom can be determined due to the final state interaction between the core hole and the valence band electrons. In addition to this, XPS is also useful for profiling the occupied DOS in the valence band. The nature of XPS allows for a probing depth of only $\approx$5 nm, due to the inelastic mean free path of excited electrons in TiO$_2$, and is thus a very surface sensitive technique.\cite{Fuentes2002} Based on the aforementioned SRIM calculations for ion implantation, the Fe content in this region is significantly less than in the bulk (1-2\% as compared to $\approx$23\% on average). In XAS, transitions are governed by dipole selection rules in which the absorption cross section is measured across a range of excitation energies. Hence, an electron is excited from a core level state to an empty valence or conduction band state and the total unoccupied DOS is probed. RXES is also governed by dipole selection rules, in this case an excitation energy is chosen to resonate with a peak in the corresponding XAS spectrum. In RXES, an incident x-ray excites an electron to produce an elementary transition in the sample, which will subsequently decay to a lower energy, with the emission of an x-ray. It is these emitted x-ray which are detected as a function of energy. By choosing an appropriate excitation energy RXES can probe specific transitions such as $d-d$, charge transfer, and even magnetic excitations.

\subsection{XPS Measurements}
The XPS survey spectra of pure and Fe-implanted TiO$_2$ (pellet and thin films) are presented in Figure \ref{Fig:XPS} (a). The samples show a relatively low carbon content, and do not contain any impurities other than Fe, as can be seen by the Fe 2$p$ signal in the doped samples.

In Figure \ref{Fig:XPS} (b) XPS Fe 2$p$-core level spectra of doped samples are shown along with spectra of reference samples FeO (Fe$^{2+}$), Fe$_2$O$_3$ (Fe$^{3+}$), and Fe-metal taken from Refs. \onlinecite{Zimmermann1999} and \onlinecite{Gao2006}. From this comparison, we see that the XPS Fe $2p$ spectrum of TiO$_2$:Fe is similar to that of Fe$_2$O$_3$, which suggests that heterovalent substitution \mbox{Fe$^{3+}$$\rightarrow$Ti$^{4+}$} takes place for both pellet and thin film materials. This was a foreseeable result because Fe on the surface of the sample will oxidize quite easily in the presence of an oxygen rich atmosphere, and XPS measurements will be quite sensitive to this within its probing depth. On the other hand, one can see the contribution of the metallic peaks in the TiO$_2$ thin film XPS signal at $\approx$707 eV and $\approx$720 eV, this is strong evidence of Fe-clustering near the surface of the thin film sample. 

XPS valence band spectra of pure and Fe doped TiO$_2$ are presented in Figure \ref{Fig:XPS} (c). O 2$s$-states are concentrated ($\approx$22.3 eV) at the bottom of the valence band, whereas O 2$p$-states prevail at the top of the valence band. Fe-doping induces Fe $3d$-states near the Fermi level as expected, as indicated by the appearance of a shoulder in the doped spectra as compared to the pure TiO$_2$ sample. For TiO$_2$:Fe (film), the Fe 3$d$-states are more pronounced and form a peak that is located at $\approx$1.5 eV, whereas in the pellet sample the Fe $3d$ peak is rather smeared. The presence of Fe 3$d$-states near the mid-gap states is consistent with the general strategy of band gap engineering TiO$_2$-based photocatalysts where the $d^0$ states are substituted by $d^n$-states. \cite{skorikov2012} But note that the presence of Fe$^{3+}$ states in the vicinity of the Fermi level is usually considered to be the main reason for the appearance of Fe$^{3+}$ induced ferromagnetism in thin film TiO$_2$:Fe.\cite{potzger2006} For completeness, it should also be stated that core level Ti $2p$ XPS spectra were investigated for all samples, but contributed very little to the conclusions posed herein. This is due to the high degree of similarity between the samples in the very weakly doped surface regions to which XPS is sensitive.

\subsection{XAS and RXES Measurements}
\begin{figure}
\includegraphics[width=3.375in]{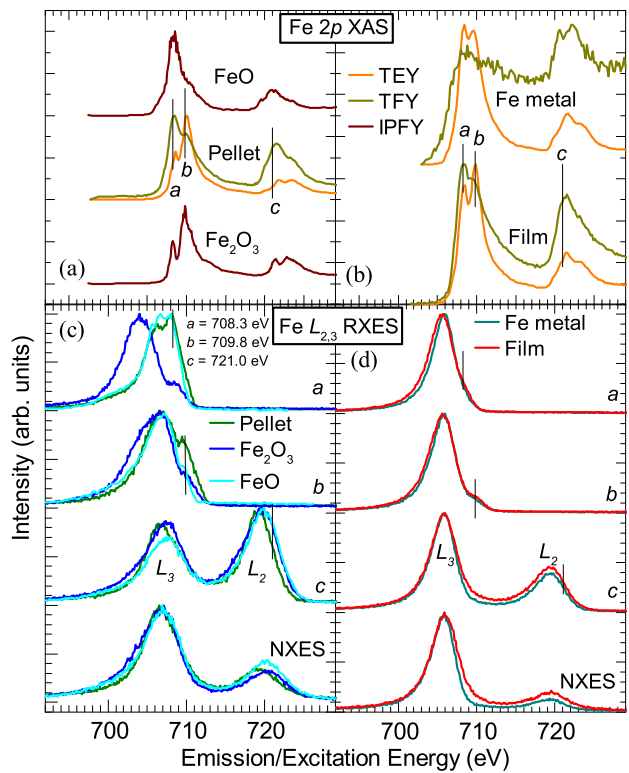}
\caption{(Color online) Comparison of Fe $2p$ XAS and XES spectra for TiO$_2$:Fe (pellet), TiO$_2$:Fe (film), Fe$_2$O$_3$, and Fe-metal. (a) The pellet sample shows a clear similarity to Fe$_2$O$_3$ and FeO in the surface sensitive TEY and bulk sensitive TFY data, respectively. (b) The thin film doped samples shows clear characteristics of Fe metal, but note that both will have suffered surface oxidation effects, and will therefore show contribitions due to the formation of Fe$^{2+}$ and Fe$^{3+}$ in what should be pure metallic Fe. (c) $L_{2,3}$ RXES spectra of the pellet sample indicates that it is indeed Fe$^{2+}$$\rightarrow$Ti$^{4+}$ substitution that occurs beneath the surface layer. This is evident at excitation energy $a$ wherein the pellet spectrum is nearly identical to the FeO spectrum, but quite different from that of Fe$_2$O$_3$. (d) Thin film RXES data is essentially identical to that of Fe metal, implying that Fe clustering also occurs beneath the surface layer.}
\label{Fig:RXES}
\end{figure}

The measurements of Fe 2$p$ XAS spectra (as shown in Figure \ref{Fig:RXES} (a,b)) show a significant difference between TiO$_2$:Fe (pellet) and TiO$_2$:Fe (film) samples. The Fe 2$p$ TEY XAS of TiO$_2$:Fe (pellet) is nearly identical to Fe$_2$O$_3$ XAS with typical multiplet splitting.\cite{crocombette1995} This is in agreement with the aforementioned XPS data which also show that the overwhelming majority of Fe near the sample surface was in the 3+ oxidation state. Quite to the contrary, the more bulk sensitive TFY XAS shows significantly more pronounced FeO characteristics. This would suggest that in the pellet sample, Fe$^{2+}$$\rightarrow$Ti$^{4+}$ substitution occurs beneath the surface where oxidizing effects are less prominent, but when exposed to oxygen, the sample soon after forms secondary phases.

On the other hand, the intensity of the first peak in the $L_3$ region of the \mbox{TiO$_2$:Fe (film)} (Figure \ref{Fig:RXES} (b)) is increased due to contributions from Fe-metal atoms as can be seen clearly when compared with the Fe metal TEY spectrum. Note that the pure Fe metal would have suffered surface oxidation effects as well, and thus it will include some contributions from FeO and Fe$_2$O$_3$, that is, the signal with not be pure Fe metal on the surface. But the marked similarity between the TEY spectra of the thin film sample and the pure Fe metal is strong evidence of primarily Fe atom clustering in the the film. The gradual `trailing off' of  the $L_2$ and $L_3$ peaks in the TFY thin film spectrum is also indicative of the metallic clustering of atoms.

In Figure \ref{Fig:RXES} (c, d), Fe $L_{2,3}$ RXES spectra for TiO$_2$:Fe (pellet) and TiO$_2$:Fe (film) are compared to reference samples. Note that the vertical lines in all spectra correspond to excitation energies $a$, $b$, and $c$ as shown in the XAS spectra above. At excitation energy $a$ ($L_3$ peak), the pellet sample is very much akin to that of the FeO reference sample, supporting the conclusions of the XAS data that it is indeed Fe$^{2+}$ substitution below the sample surface.

The TiO$_2$ thin film RXES spectra in Figure \ref{Fig:RXES} (d) nearly exactly reproduce the Fe metal spectra at all excitation energies. This is convincing evidence that Fe atoms aggregate, and along with the XPS and XAS data, it can be firmly concluded that in the thin film there is an overwhelming tendency for the Fe atoms to cluster together under the conditions of sample fabrication presented herein.

The relative I($L_2$)/I($L_3$) intensity ratio of the TiO$_2$:Fe (film) sample excited at the $L_2$ threshold (point $c$ at 721.0 eV) is much smaller than that of the TiO$_2$:Fe (pellet) sample. This intensity ratio is similar to that of FeO/Fe$_2$O$_3$ and \mbox{Fe metal (Figure \ref{Fig:RXES}).}  The I($L_2$)/I($L_3$) intensity ratio is usually related to the probability of non-radiative $L_2$$L_3$$M_{4,5}$ Coster-Kronig (C$-$K) transitions (in which an electron transitions from within the same shell, for example $2p \rightarrow 2s$), and also to the ratio of total photo absorption coefficients ($\mu_3/\mu_2$) for excitation energies at the $L_2$ and $L_3$ absorption threshold.\cite{kurmaev2005} Since the ratio of total photon absorption coefficients depends only on the excitation energy, the I($L_2$)/I($L_3$) intensity ratio of RXES spectra taken at the same excitation energy is determined by the C$-$K transitions alone, which are in turn governed by the number of free $d$-electrons around the target atom. The I($L_2$)/I($L_3$) ratio of Fe atoms in the mixed \mbox{Fe$^{3+}$ $+$ Fe$^0$} state (as in the film) should be suppressed in comparison with Fe atoms in the Fe$^{2+}$ $+$ Fe$^{3+}$ state (as in the pellet sample). Therefore, Fe $L_2$ RXES measurements confirm the existence of Fe$^{3+}$ species in both pellet and thin film TiO$_2$:Fe samples. In addition to this, there is strong suppression of the I($L_2$)/I($L_3$) ratio in the thin film sample due to the additional clustering of Fe-atoms (i.e. Fe$^0$ atoms have more free $d$-electrons and therefore a suppressed I($L_2$)/I($L_3$) ratio). Both RXES and XAS spectra indicate the clustering of Fe-atoms in TiO$_2$:Fe (film), which is in good agreement with the more surface sensitive XPS Fe 2$p$-measurements (Figure \ref{Fig:XPS}).

\begin{figure}
\includegraphics[width=3.375in]{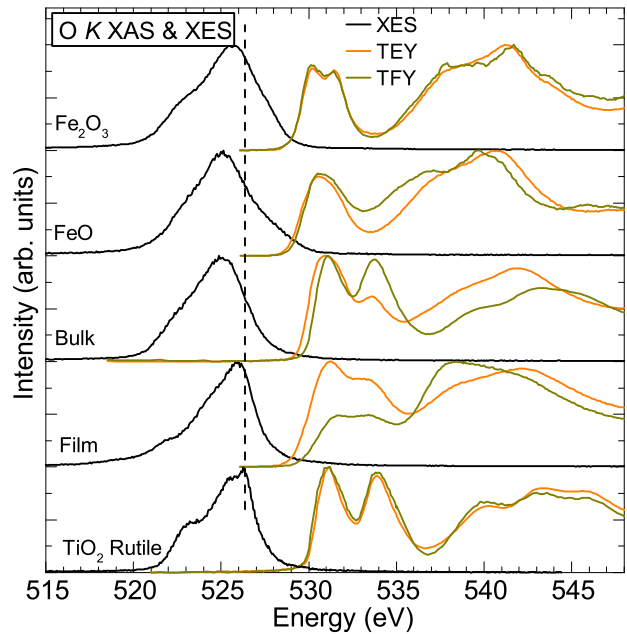}
\caption{(Color online) O $K$ XAS and XES. Doping of TiO$_2$ with Fe in pellet and thin film indicate that there are contributions from secondary phases due to oxidation on the surface of both samples. This is clear from the large suppression of the second main peak in the surface sensitive TEY data, in the same energy region in which FeO and Fe$_2$O$_3$ show a weak signal. TFY spectra largely probe beneath the implantation depth, and is therefore very much like pure TiO$_2$ in the pellet sample, and show evidence of the Si substrate in the thin film. XES data suggests that Fe doping shifts the main O peak to lower energies, as in FeO and Fe$_2$O$_3$ reference samples.}
\label{fig:oxygen}
\end{figure}

Figure \ref{fig:oxygen} shows XAS and non-resonant XES oxygen $K$-edge spectra for both pellet and thin film Fe doped TiO$_2$ samples, pure TiO$_2$, and reference samples FeO and Fe$_2$O$_3$. Oxygen XES spectra vary between the samples due to the differing crystal structures, and the Fe-coordination level. The shift of the main peak in the XES data of the doped samples to lower energies can be attributed to Fe doping. This can be seen by looking at the main O peak in FeO and Fe$_2$O$_3$ and noting that they are noticeably lower than that of pure TiO$_2$.

The O $K$-edge TFY spectra of the pellet sample is nearly identical to that of the pure TiO$_2$ sample because, as noted earlier, the maximum depth of implantation of Fe ions was 45 nm, while TFY probes on the order of \mbox{$\sim$150-200 nm}. This means that fluorescence yield is largely obtained from the pure TiO$_2$, beneath the level of ion implantation (likely less than 25\% of the signal is from the actual implantation region). Whereas the film TFY data show a significant suppression of the first two peaks in comparison with pure TiO$_2$. This is likely due to an oxide layer of SiO$_2$ forming at the substrate-film boundary, and it is this O $K$ signal showing through the TiO$_2$. This is clear if one compares it to the onset of the O $K$-edge of SiO$_2$ in Ref. \onlinecite{Zatsepin2011}, in which there are no electronic states in the suppressed region.

The surface sensitive TEY data show clear shifts to lower energies of the onset of the O $K$-edge (similar to that of reference samples), again indicating the expected formation of secondary phases on the sample surfaces. This shifting to lower energies of the onset of the conduction band signifies that Fe doping decreases the band gap on the sample surface, in agreement with the XPS VB data of Figure \ref{Fig:XPS}. The onset of O $K$-edge XAS is very close to the true ground state conduction band, this is because the O $1s$ core hole only leads to a very minor perturbation on the energy of the unoccupied states.\cite{McLeod2012} This is in contrast to the Fe $L$-edge spectra (which would offer the most direct probe of the electronic structure in the vicinity of the dopants), because the Fe $2p$ core hole will considerably perturb the onset of the $L$-edge from the true ground state conduction band. 

The TEY spectra of the doped samples also show a large suppression of the second main peak as compared to the pure TiO$_2$ sample. This is in the same energy range as the large dips in intensity of the reference samples, again indicating that secondary phases of Fe are prevalent in the surface region.


\subsection{DFT Calculations}
The x-ray spectroscopy measurements shown in the previous section indicates that Fe-doping in the pellet TiO$_2$ rutile sample induces Fe$^{2+}$ substitution at Ti$^{4+}$ sites beneath the surface layer; and that Fe$^{3+}$ substitution occurs near the surface (note that both XAS in TEY mode and XPS have a probe depth of only $\sim$5nm). Therefore, we have spectroscopic data to test the validity of our DFT structural optimizations and electronic structure calculations, these calculations can then be used to obtain a better understanding of the influence of Fe-doping in TiO$_2$.


The calculations of the formation energy for iron impurities in both anatase and rutile TiO$_2$ are shown in Table \ref{tbl:table1}. As the criterion for the observed oxidation states, it is possible to map a given impurity to the observed phase of the implanted Fe atoms. A single substitutional ($S$) impurity (i.e. an Fe atom replacing a Ti atom) corresponds to FeO$_2$. This is a very unstable 4+ oxidation state for iron, and it is therefore not surprising that it is not observed in our experimental data, and not favoured in the calculations. Similarly, an $S+I$ (substitution and nearby interstitial Fe atom) corresponds to a 2+ oxidation state, due to the favourable formation of local Fe-O bonding in the form of Fe($S$)O and Fe($I$)O. And $2S+I$ can be regarded as the favourable formation of a combination of Fe$_2$O$_3$ (Fe$^{3+}$) and Fe$_3$O$_4$ (both Fe$^{2+}$ and Fe$^{3+}$), in the form of $2$Fe($S$)Fe($I$)O$4$. Note that this mapping provides a crude, but logical way of relating the DFT calculations to the observed valencies; and the actual formation of a given oxidation state or phase of iron will be a complex process depending upon many factors that we cannot explicitly account for. Nevertheless, the calculations provide valuable information in understanding the cause of the observed oxidation states.

The first thing to notice is the near identical formation energies of all defects on the surfaces of both samples. This is in agreement with our highly surface sensitive spectroscopic XPS and TEY data, both of which indicated a strong Fe$^{3+}$ signal. But consider that the formation energies for $S+I$ and $2S+I$ on the surface of anatase are more closely spaced, and less in energy as compared to the rutile sample (0.62 eV and 0.92 eV compared to 0.68 eV and 1.01 eV, respectively). This is likely the cause of a largely metallic Fe signal on the surface of our anatase sample, as it could lead to the formation of several nearby $S+I$ and $2S+I$ defects. The formation of  neighbouring $2S+I$ and $S+I$ impurities can generally be considered as the aggregation of Fe atoms, and it is this aggregation that is observed as a metallic signal in our data. The reason for this is because in anatase TiO$_2$ the distance between impurity atoms in cation (substitutional) sites and interstitial sites is \mbox{$\sim$2.67 \AA}, and in pure Fe metal the atomic spacing is \mbox{$\sim$2.47 \AA}. That is, the distance between substitutional and interstitial sites is comparable to the distance between metallic Fe atoms. We can therefore conclude that an arrangement of  adjacent $S+I$ and $2S+I$ Fe-impurities can be related to the formation of iron clusters in TiO$_2$:Fe thin films, in agreement with the observations in XPS, XAS, and XES measurements. This is in contrast to the ion implantation in SiO$_2$ and ZnO  thin film hosts, wherein substitution below the surface layer was energetically favourable and the clustering of metallic atoms was not observed (see our previous work regarding Pb and Sn in SiO$_2$ \cite{Green2012} and Fe in ZnO\cite{Mcleod}).  It was also calculated that $3+$ substitutional iron impurities in TiO$_2$ have a magnetic moment of 3.21 $\mu$B for single impurities, and 3.28 $\mu$B for a pair of nearest neighbours. 


\begin{table}
\centering
\caption{The calculated formation energies (in eV) for 3$d$-impurity atoms for substitutional ($S$)  impurities and their various combinations with interstitial ($I$) defects (\mbox{$S+I$} and \mbox{$2S+I$)} in TiO$_2$ rutile and antase phases.}
\begin{tabular}{ l  c  c  c  c  }
    \hline \hline
    Sample & Dopant Location & $S$ & $S+I$ & $2S+I$\\ \hline
    Anatase & Surface & 1.58 & 0.62 & 0.92 \\ 
                  & Bulk & 2.09 & 1.59 & 0.67\\ \hline
    Rutile & Surface & 1.58 & 0.68 & 1.01 \\ 
                  & Bulk & 1.86 & 0.72 & 0.88\\ \hline
\end{tabular}
\label{tbl:table1}
\end{table}


DFT calculations for rutile show that $S+I$ (and therefore Fe$^{2+}$) impurities are the most favourable configuration (0.72 eV) below the surface region, this is in agreement with the spectroscopic data of the previous sections. On the other hand, it was calculated that $2S+I$ impurities come at only a slightly higher energy cost (0.88 eV), and thus we may expect to see an Fe$^{3+}$ signal in our pellet sample (which we do not). To explain this we argue that the formation of Fe$^{2+}$ substitutional impurities in the pellet sample can be justified in terms of the thin film versus pellet geometry, rather than the differences between rutile and anatase crystal phases. Our main motivation for this claim is that the polycrystalline thin film was synthesized with crystallites of $\sim$5 nm in size, while in the pellet sample the crystallites were $\sim$200 nm. The surface to volume ratio scales as $1/r$, and therefore there is a large difference in effective surface to volume ratio inherent in the samples ($200/5 = 40$ times the effective surface area in the thin film). If we assume oxygen can only reach the surface or penetrate between crystallites, we can then conclude that there is a significant relative deficiency of oxygen in the pellet samples, and thus is highly conducive to the formation of Fe$^{2+}$ as opposed to Fe$^{3+}$.

For conventional transition metal doped semiconductors, the observed configurations of impurities and their relation to the methods of samples fabrication (sol-gel, molecular beam epitaxy, etc.) can be explained by the kinetics of the material's formation.\cite{chang2007} But in our case, the insertion of impurities occurs \emph{after} the fabrication of the samples, this is in contrast to previously used methods wherein impurities were introduced to the host during the fabrication process.\cite{Nagaveni2004, Mangham2011, Zhang2003, Zhang2006} 

Differences between the configurations of impurities in pellet and thin film samples has previously been discussed in terms of different surface to volume ratios.\cite{Mcleod} To examine the effect on the surface, we have calculated formation energies for $1S$ and $1S+1I$ defects on the hydrogen passivated surface of TiO$_2$. The obtained energies for these two defects were +0.17 eV/Fe and -0.97 eV/Fe, respectively. We can therefore say that the formation of a $1S+1I$ defect is more energetically favourable in the vicinity of the surface than in the bulk as compared to only substitution ($1S$). Therefore, ultimately we can explain different amounts of Fe clustering in pellet and thin films samples in terms of different surface to volume ratios.

\section{Conclusion}
To conclude, we have studied the formation of structural defects induced by Fe-ion implantation of TiO$_2$ pellet and thin films with XPS, XAS and RXES analytical techniques. The results were compared with DFT calculations of the formation energies for different configurations of structural defects. It was found that Fe$^{2+}$$\rightarrow$Ti$^{4+}$ heterovalent substitution takes place in pellet TiO$_2$:Fe samples below the surface layer, which is prone to oxidation effects and thus Fe$^{3+}$ is largely detected on the sample surface. In thin film TiO$_2$:Fe samples, in addition to Fe$^{3+}$ on the surface, primarily the clustering of Fe-atoms was observed. This suggests that under Fe-ion implantation, the controlled and reproducible data can be obtained only for the pellet TiO$_2$:Fe material.

\begin{acknowledgments}
We gratefully acknowledge support from the Natural Sciences and Engineering Research Council of Canada (NSERC) and the Canada Research Chair program. This work was done with partial support of Ural Division of Russian Academy of Sciences (Project 12-I-2-2040), Ministry of Science and Education of Russian Federation (GK No.16.513.11.3007) and Russian Foundation for Basic Research (Project 13-08-00059).
\end{acknowledgments}


%

\end{document}